\input epsf
\documentstyle{mn}
\newif\ifAMStwofonts

\def\lesssim{\mathrel{\hbox{\rlap{\hbox{\lower4pt\hbox{$\sim$}}}\hbox{$<$}}}}
\def\gtrsim{\mathrel{\hbox{\rlap{\hbox{\lower4pt\hbox{$\sim$}}}\hbox{$>$}}}}
\def\apj{ApJ}
\def\apjs{ApJS}
\def\aj{AJ}
\def\aap{A\&\hskip-1pt A}

\def\mnras{MNRAS}

\newcommand{\br}{\mbox{\boldmath $r$}}
\def\eqalign#1{\null\,\vcenter{\openup\jot 
   \ialign{\strut\hfil$\displaystyle{##}$&$
      \displaystyle{{}##}$\hfil \crcr#1\crcr}}\,}

\title[Microlensing Color Change Measured by the DIA Method]
      {\LARGE {\bf Colour Change Measurements of Gravitational Microlensing \\
       Events by Using the Difference Image Analysis Method}}

\author[Han \& Park]
       {Cheongho Han and Seong-Hong Park\\
        Department of Astronomy \& Space Science, \\
        Chungbuk National University, \\
 	Chongju, Korea 361-763, \\
	{\tt cheongho,parksh@astro.chungbuk.ac.kr}}

\date{Accepted:\\
      Received: }

\pagerange{\pageref{firstpage}--\pageref{lastpage}}

\begin{document}

\maketitle

\label{firstpage}

\begin{abstract}
Detecting colour changes of a gravitational microlensing event induced
by the limb-darkened extended source effect is important to obtain useful
information both about the lens and source star.  However, precise
measurements of the colour changes are hampered by blending, which also
causes colour changes of the event.  In this paper, we show that although
the colour change measured from the subtracted image by using the recently
developed photometric method of the ``difference image analysis'' (DIA)
differs from the colour change measured by using the conventional method
based on the extraction of the individual source stars' point spread
functions, the curve of the colour changes (colour curve) constructed by
using the DIA method enables one to obtain the same information about the
lens and source star, but with significantly reduced uncertainties due to the
absence of blending.  We investigate the patterns of the DIA colour curves
for both single lens and binary lens events by constructing colour
change maps.
\end{abstract}


\section{Introduction}

The amplification of source flux caused by gravitational lensing can
become theoretically infinite.  Points in the source plane on which the
amplification of a point source becomes infinite are called caustics.
For a point lens, the caustic is a single point behind the lens.  For a
binary lens, the number of caustics is multiple and they form closed
curves.  For a real microlensing event, however, the source star has
an extended size, and thus the observed amplification  deviates from
that of a point source event and becomes always finite: extended source
effect (Schneider, Ehlers, \& Falco 1992; Witt \& Mao 1994).  The
deviation of the light curve caused by the extended source effect
becomes important as the source passes very close to the lens caustics.

Detection of the extended source effect is important because one can
obtain useful information both about the lens and source star.  First,
a caustic crossing event provides an opportunity to measure how long it
takes for the caustics to transit the face of the source star.  By using
the source radius crossing time $t_\ast$, along with an independent
determination of the source star size $\theta_\ast$, one can determine
the lens proper motion with respect to the source star by $\mu=\theta_\ast
/t_\ast$ (Witt \& Mao 1994; Gould 1994; Maoz \& Gould 1994; Nemiroff \&
Wickramasinghe 1994; Loeb \& Sasselov 1995; Alcock et al.\ 1997b, 1997c,
1999a; Afonso et al.\ 1998; Udalski et al.\ 1998; Albrow et al.\ 1999a).
Once the value of $\mu$ is determined, the mass and location of the lens
can be significantly better determined.  Second, by analyzing the light
curve of an event with the source trajectory passing very close to the
caustics, one can investigate the surface structure of the source star
such as the surface intensity profile and spots (Valls-Gabaud 1994,
1998; Sasselov 1997; Gaudi \& Gould 1999; Heyrovsk\'y \& Sasselov 2000;
Albrow et al.\ 1999b; Han et al.\ 2000).

In addition to detecting deviations in the light curve, the extended
source effect can also be detected by measuring the colour changes
of the source star during the event.  The colour changes occur due to
the differential amplification over the limb-darkened source star surface
(see \S\ 3.1).  Throughout this paper, we use a term `colour curve'
to refer to the colour changes of an event induced by the extended
source effect as a function of time.  Detection of the extended source
effect from the colour measurements has the following advantages compared
to the detection from a single band photometry.  First, by detecting
colour changes one can increase the chance to detect the extended
source effect.  If the source star approaches very close to the lens
caustics but does not transit the caustics, the amplification induced
by the extended source effect simply masquerades as changes in lensing
parameters, and thus cannot be detected.  By contrast, the colour
change cannot be mimicked by the change in lensing parameters, because
a point source lensing event should be achromatic (Gould \& Welch 1996).
Second, by measuring the colour changes one can determine the lens
proper motion with relative ease.  This is because one can measure
$t_\ast$, and thus $\mu$, by simply measuring the turning time of the
colour curve (Han, Park, \& Jeong 2000, see \S\ 4 for more details).
To determine $t_\ast$ from the light curve, on the other hand, it is
required to fit the overall light curve.

However, precise measurements of the colour changes induced by the extended
source effect is hampered by blending, which also causes colour changes of
the event (Kamionkowski 1995; Buchalter, Kamionkowski, \& Rich 1996).
To increase the event rate, the current lensing experiments are being
conducted towards very dense star fields such as the Galactic bulge
and the Magellanic Clouds (Alcock et al.\ 1993; Aubourg et al.\ 1993;
Udalski et al.\ 1993; Alard \& Guibert 1997).  When the brightness of
a source star located towards these very dense star fields is measured
by using the  conventional method based on the extraction of the individual
source stars' point spread functions, it is very likely that the
measured source star flux is affected by the unwanted light from nearby
unresolved stars (Di Stefano \& Esin 1995; Alard 1997; Alcock et al.\
1997a; Palanque-Delabrouille et al.\ 1998; Wo\'zniak \& Paczy\'nski 1997;
Han 1999; Han, Jeong, \& Kim 1998).  Since blended stars in general
have different colours from the colour of the lensed source star, the
measured colour is affected by blending.  Han, Park, \& Jeong\ (2000)
pointed out the seriousness of blending in colour change measurements by
demonstrating that even for a very small fraction ($\lesssim 2\%$) of
blended light, the colour change caused by blending can be equivalent
to the colour change induced by the extended source effect.  Therefore,
for the precise measurements of the colour changes induced by the
extended source effect, it will be essential to correct for or remove
the blending effect.

Although numerous methods have been proposed for the correction of
blending (Alard, Mao, \& Guibert 1995; Alard 1996; Han 1997, 1998;
Goldberg 1998; Goldberg \& Wo\'zniak 1998; Han \& Kim 1999), most
of these methods either have limited applicabilities only to several
special cases of blended events or are less practical due to the
requirement of space observations.\footnote{We note, however, that
the MACHO group had {\it Hubble Space Telescope} followup observations
of the source star for one of the events they have detected to
clearly identify the lensed source star.  From these observations,
they could accurately estimate the lens mass by correcting blending
(NASA press release \#STScI-PR00-03).}  On the other hand, with the
recently developed photometric technique of the ``difference image
analysis'' (DIA) method, one can measure the blending-free light
variations of general microlensing events (Tomaney \& Crotts 1996;
Alard \& Lupton 1998; Alard 1999; Alcock et al.\ 1999b, 1999c).
The DIA method detects and measures the variation of the source
star flux by subtracting an observed image from a convolved and
normalized reference image.  Since the blended light is subtracted
during the image subtraction process, one can improve the photometric
precision by removing the effect of blending.  At the moment most
microlensing experiments do not use the DIA method in their data analysis.
Using the DIA method would require that either the whole microlensing
survey process should switch to the DIA method, or that systematic
re-processing of the light curves should be performed once events have
been detected using the conventional photometric method.

Due to the methodological difference of the DIA photometry from the
conventional photometry, the colour change of an event measured by
the using the DIA method (DIA colour change) differs from that measured
by the conventional method (see \S\ 3).  In this paper, we show, however,
that the DIA colour curve of an event enables one to obtain the same
information about the lens and source star, but with significantly
reduced uncertainties due to the absence of blending.  We investigate
the patterns of the DIA colour curves for both single and binary lens
events by constructing colour change maps.

\section{Gravitational Amplification}

\subsection{Point Source Events}

If a microlensing event with a point source is caused by a single lens,
the light curve of the event is represented by
\begin{equation}
A_{\rm p} = {u^2+2\over u(u^2+4)^{1/2}};\qquad
u = \left[ \left( {t-t_0\over t_{\rm E}}\right)^2 + \beta^2 \right]^{1/2},
\end{equation}
where the subscript `p' denotes the event is occurred to a point source,
$u$ represents the lens-source separation normalized by the angular
Einstein ring radius $\theta_{\rm E}$, and the lensing parameters $\beta$,
$t_0$, and $t_{\rm E}$ represent the impact parameter of the lens-source
encounter, the time of maximum amplification, and the Einstein ring radius
crossing time (Einstein time scale), respectively.  The angular Einstein
ring represents the effective region of gravitational amplification and
its size is related to the physical parameters of the lens by
\begin{equation}
\theta_{\rm E}= \left( 
{4GM\over c^2} {D_{ls}\over D_{ol}D_{os}}
\right)^{1/2},
\end{equation}
where $M$ is the mass of the lens and $D_{ol}$, $D_{ls}$, and $D_{os}$
represent the separations between the observer, the lens, and the source,
respectively. When the source is located within the Einstein ring, the
source star amplification becomes greater than $3/\sqrt{5}$.

If an event is caused by a binary lens, on the other hand, the resulting
light curve differs from that of a single lens event.  When lengths are
normalized by the combined angular Einstein ring radius, which represents
the angular Einstein ring radius with a lens mass equal to the total mass
of the binary system, the lens equation in complex notation for the binary
lens system is represented by
\begin{equation}
\zeta = z + {m_{1} \over \bar{z}_{1}-\bar{z}} 
          + {m_{2} \over \bar{z}_{2}-\bar{z}},
\end{equation}
where $m_1$ and $m_2$ are the mass fractions of individual lenses (and
thus $m_1+m_2=1$), $z_1$ and $z_2$ are the positions of the lenses,
$\zeta = \xi +i\eta$ and $z=x+iy$ are the positions of the source and
images, and $\bar{z}$ denotes the complex conjugate of $z$ (Witt 1990).
The amplification of each image, $A_{{\rm p},i}$, is given by the Jacobian
of the transformation (3) evaluated at the images position, i.e.\
\begin{equation}
A_{{\rm p},i} = \left({1\over \vert {\rm det}\ J\vert} \right)_{z=z_i};
\qquad {\rm det}\ J = 1-{\partial\zeta\over\partial\bar{z}}
{\overline{\partial\zeta}\over\partial\bar{z}}.
\end{equation}
Then the total amplification of the event is given by the sum of the
individual amplifications, i.e.\ $A_{\rm p}=\sum_i A_{{\rm p},i}$.
The caustic refers to the source position on which the total amplification
becomes infinite, i.e. ${\rm det}\ J=0$.

\subsection{Extended Source Events}

If the source of an event has an extended size, the light curve of the
event deviates from that of a point source event.  For both single and 
binary lens events, the amplification of an extended source event is the 
weighted mean of the amplification factor over the source star disk, i.e.\
\begin{equation}
A_\nu = {
\int_{0}^{r_\ast} I_\nu(r)A_{\rm p}(\left\vert \br-\br_{L}
\right\vert)rdr \over \int_{0}^{r_\ast} I_\nu(r)rdr },
\end{equation}
where $I_\nu(r)$ is the radial surface intensity distribution of the
source star with a radius $r_\ast$ and the vector $\br_{L}$ and
$\br$ represent the displacement vector of the source star center
with respect to the lens and the orientation vector of a point on
the source star surface with respect to the center of the source star,
respectively.

\section{Colour Changes}
\subsection{Measured by the Conventional Photometric Method}

In addition to causing deviations in the light curve from that of a point
source event, the extended source effect makes the gravitational amplification
become wavelength dependent, causing colour changes during the event.  The
colour changes occur due to the wavelength dependency of the source star
radial surface intensity profile caused by limb darkening.  If we define
$F_{0,\nu i}=2\pi \int_0^{r_\ast} I_{\nu i} (r) rdr$, where $ i=1,\ 2$, and
$m_{\nu i}$ as the unamplified source star fluxes and the corresponding
magnitudes measured at two different wavelength pass-bands $\nu 1$ and
$\nu 2$, the colour of the {\it un-blended} source star measured at a time $t$ 
by using the conventional method is computed by
\begin{equation}
(m_{\nu 2}-m_{\nu 1})_0 (t)=
-2.5\log 
\left[{A_{\nu 2}(t) F_{0,\nu 2}\over A_{\nu 1}(t) F_{0,\nu 1}}\right].
\end{equation}
Then the colour curve of the event is represented by
\begin{equation}
\eqalign{
\Delta (m_{\nu 2}-m_{\nu 1})_0 (t) & =  \cr
			           & \hskip-10pt -2.5 \log
\left\{ \left[ {A_{\nu 2}(t) \over A_{\nu 1}(t)}\right]
\left[ {A_{\nu 2}(t_{\rm ref}) \over A_{\nu 1}(t_{\rm ref})}\right]^{-1}
\right\},  \cr
}
\end{equation}
where $t_{\rm ref}$ represents the reference time for the colour change
measurements.  For a point source event, $A_{\nu 1}(t)=A_{\nu 2}(t)$ (i.e.\
achromatic), and thus $(m_{\nu 2}- m_{\nu 1})_0= -2.5\log (F_{0,\nu 2}/ 
F_{0,\nu 1})$, implying that the colour of the source during the event
equals to that of the un-amplified source.  For a limb-darkened extended
source event, on the other hand, as the source passes close to caustics,
different parts of the source star disk with varying surface intensity
and spectral energy distribution are amplified by different amount due
to the differences in distance to the lens.  As a result, the amplification
becomes wavelength dependent, i.e.\ $A_{\nu 1}(t)\neq A_{\nu 2}(t)$,
and the colour changes during the event.  Once the colour curve of the
event is constructed, the lens proper motion and the source star surface
intensity profile are determined by statistically comparing the observed
colour curve to the theoretical curves with various models of limb
darkening and source star size.

However, precise measurement of the colour changes induced by the extended
source effect is hampered by blending, which is another mechanism causing
colour changes of microlensing events.  If one defines $B_{\nu i}$ as the
blended amount of flux in two passbands, the measured colour of a blended
event becomes
\begin{equation}
\eqalign{
(m_{\nu 2}-m_{\nu 1})(t) & = -2.5\log \left[{A_{\nu 2}(t)F_{0,\nu 2}+B_{\nu 2}\over 
                              A_{\nu 1}(t)F_{0,\nu 1}+B_{\nu 1}}\right] \cr
                         & = -2.5\log \left[ {A_{\nu 2}(t)+f_{\nu 2}
                             \over A_{\nu 1}(t)+f_{\nu 1}}\right], \cr
}
\end{equation}
where $f_{\nu i}=B_{\nu i}/F_{0,\nu i}$ represent the ratios between the
blended and the baseline source flux in the individual pass-bands.  Then
the colour curve of a {\it blended} event is represented by
\begin{equation}
\eqalign{
\Delta(m_{\nu 2}-m_{\nu 1}) = &  \cr  
  & \hskip-60pt 
-2.5\log \left\{
\left[ {A_{\nu 2}(t)+f_{\nu 2} \over A_{\nu 1}(t)+f_{\nu 1}}\right] \
\left[ {A_{\nu 2}(t_{\rm ref})+f_{\nu 2} \over A_{\nu 1}(t_{\rm ref})+
f_{\nu 1}}\right]^{-1} \right\}. \cr
}
\end{equation}
One finds that equation (9) includes two additional blending parameters
of $f_{\nu 1}$ and $f_{\nu 2}$ compared to the un-blended event colour
curve in equation (7).  Therefore, to determine the lens proper motion
and the source star surface intensity profile from the fit of the blended
event colour curve, it is required to include these additional parameters,
causing increased uncertainties in the determined quantities.

\subsection{Measured by the DIA Method}
The colour changes induced by the extended source effect can also be
measured by using the DIA method.  The flux of a source star measured
from the subtracted image by using the DIA method is represented by
\begin{equation}
F_\nu=F_{\nu,{\rm obs}}-F_{\nu, {\rm ref}} = (A_\nu-1) F_{\nu,0},
\end{equation}
where $F_{\nu,{\rm obs}}=A_\nu F_{0,\nu} + B_\nu$ and $F_{0,{\rm ref}}
= F_{0,\nu}+B_\nu$ represent the source star fluxes measured from the
images obtained during the progress of the event and from the reference
image, respectively.  Then the DIA colour curve of the event is
represented by
\begin{equation}
\eqalign{
\Delta(m_{\nu 2}-m_{\nu 1})_{\rm DIA} = &   \cr
   &  \hskip-60pt
-2.5 \log \left\{ \left[ { A_{\nu 2}(t)-1\over A_{\nu 1}(t)-1}\right]
\left[ { A_{\nu 2}(t_{\rm ref})-1\over A_{\nu 1}(t_{\rm ref})-1}
\right]^{-1} \right\}. \cr
}
\end{equation}
From equation (11), one finds that the DIA colour curve does not depends
on the blending parameters, and thus it is free from blending. One also
finds that although the DIA colour curve has a different form from the
curve of the un-blended event constructed by using the conventional method
[cf.\ equation (7)], both curves depend on the same parameters of
$A_{\nu 1}$ and $A_{\nu 2}$.  Therefore, from the DIA colour curve
one can obtain the same information about the lens and source star as
that obtained from the colour curve constructed by using the conventional
method, but with significantly reduced uncertainties due to the absence
of blending.

\begin{figure}
\epsfysize=14.5cm
\centerline{\epsfbox{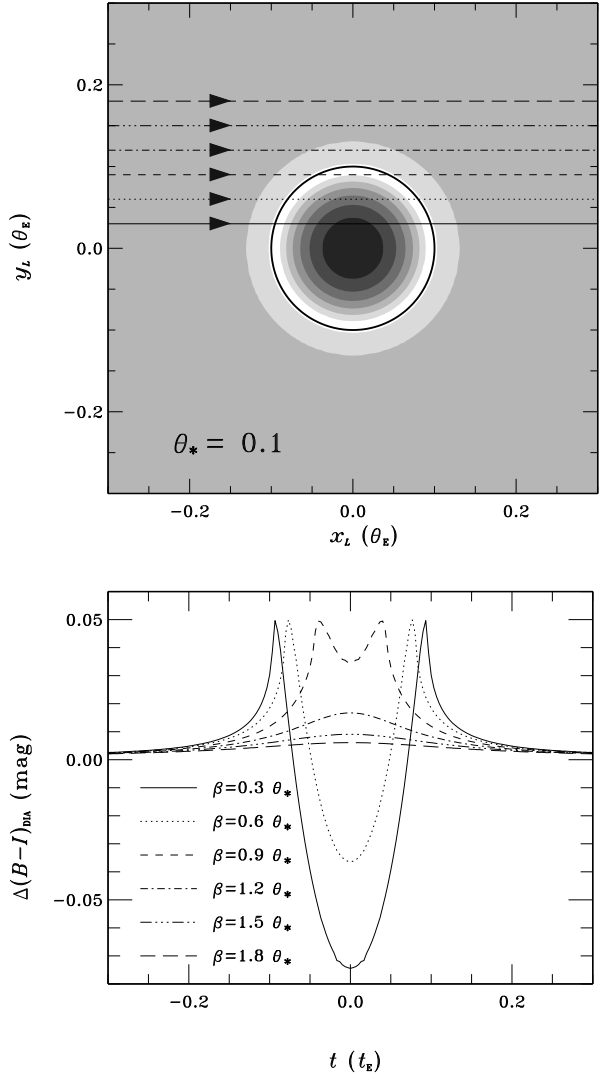}}
\vskip-0.3cm
\caption{
The colour map (upper panel) and curves (lower panel) for a single lens
system.  The map is presented as a function of lens position $(x_L, y_L)$
with respect to the source star (the thick solid circle).  The source star
has an angular radius (normalized by $\theta_{\rm E}$) of $\theta_\ast=0.1$.
The reference of the colour change is chosen so that $\Delta (B-I)_{\rm DIA}=0$
when the source star is not gravitationally amplified.  Grey scale becomes
brighter (darker) as the colour of the source star becomes redder (bluer),
and the tone of the grey scale changes with a colour change interval of
$0.02$ mag from the darkest region where $\Delta (B-I)_{\rm DIA}\leq -0.06$
mag to the brightest region where $\Delta (B-I)_{\rm DIA} > 0.04$.  The lens 
trajectories (the straight lines in the upper panel) and their corresponding 
colour curves are drawn by the same line types.
}
\end{figure}

\section{Patterns of DIA Colour Change Curves}

In this section, we investigate the patterns of the DIA colour curves
for both single and binary lens events.  To see the patterns, we construct
the colour maps, which represent the colour changes as a function of source
(lens) position with respect to the position of the lens (source).  The
constructed map are presented in the upper panel of Figure 1 for the
single lens system and Figure 2 for the binary lens system.  For the
construction of the maps, we assume that the source star has an angular
radius (normalized by $\theta_{\rm E}$) of $\theta_\ast=0.1$ and it is
observed in $B$ and $I$ bands.  The un-blended colour of the source star
before amplification is $(B-I)_0=2.15$ mag, which corresponds to that of
a K-type giant (Allen 1973; Schmidt-Kaler 1982; Peletier 1989). For the
surface intensity profile, we adopt a linear form of
\begin{equation}
I_{\nu}(r) = 1-{\cal C}_\nu \left[ 1-\sqrt{1-(r/r_\ast)^2}\right],
\end{equation}
where the limb-darkening coefficients are ${\cal C}_{B}=0.912$ and
${\cal C}_{I}=0.053$, respectively, which correspond to those of a K giant
with $T_{\rm eff}=4,750$ K, $\log g= 2.0$, and a metallicity similar to
the sun (Van Hamme 1993).  For the binary lens system, we adopt a binary
separation (also normalized by $\theta_{\rm E}$) of $\ell=1.0$ and a mass
ratio of $q=m_1/m_2=1.0$.  The single lens map is presented as a function
of lens positions $(x_L,y_L)$ with respect the source star (the thick solid
circle with its center at the origin), while the binary lens map is
presented as a function of source position $(\xi,\eta)$ with respect to
the center of the binary system.  All lengths in the maps are normalized
by $\theta_{\rm E}$.  For both maps, we choose the reference of the colour
change measurements to be $\Delta (B-I)_{\rm DIA}=0$ when the source star
is not gravitationally amplified.  Grey scale is drawn so that it becomes
brighter (darker) as the colour of the source star becomes redder (bluer),
and the tone of the grey scale changes for every colour change of $0.02$
mag.  For the single lens map, the grey tone change from the darkest
region where $\Delta (B-I)_{\rm DIA}\leq -0.06$ mag to the brightest region
where $\Delta (B-I)_{\rm DIA} > 0.04$ mag.  For the binary lens map,
the tone changes from the darkest region where $\Delta (B-I)_{\rm DIA}\leq 
-0.05$ mag to the brightest region where $\Delta (B-I)_{\rm DIA} > 0.07$
mag.  To better show the fine structures of the binary lens event map in
the region near the caustics (marked by thick solid curves), the region
enclosed by a box (drawn by a short-dashed line) is expanded and presented
in the upper panel of Figure 3.  The straight lines in the upper panels of
Figure 1 and 3 represent various lens (for the single lens case) or source
(for the binary lens case) trajectories and the colour curves resulting
from the trajectories are presented in the lower panels of the individual
figures.  The line types of the colour curves are selected so that they
match with those of the corresponding lens or source trajectories.

\begin{figure*}
\epsfysize=16cm
\centerline{\epsfbox{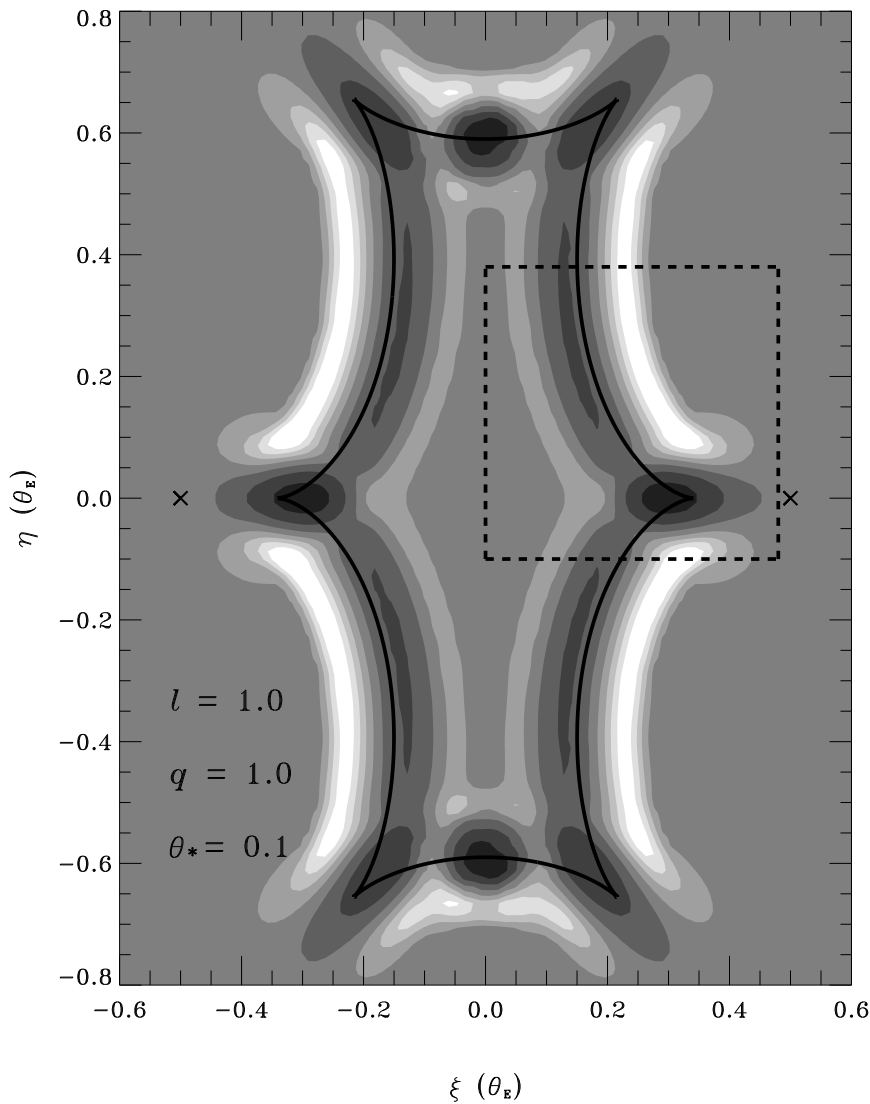}}
\vskip-1.4cm
\caption{
The colour map for a binary lens system.  The map is presented as a function
of source star position $(\xi,\eta)$ with respect to the positions of the
binary lenses (marked by `x').  The source star has an angular radius
$\theta_\ast=0.1$.  The binary system has a separation (also normalized by
$\theta_{\rm E}$) between its components of $\ell=1.0$ and a mass ratio of
$q=1.0$.  The closed figure drawn by a thick solid curve represents the
caustics.  Grey scale is drawn by the same way as in Figure 1, but its tone
changes from the darkest region where $\Delta (B-I)_{\rm DIA} \leq -0.05$
mag to the brightest region where $\Delta (B-I)_{\rm DIA} > 0.07$ mag.
To better show the fine structures of the map in a region near the caustics,
the region enclosed by a box is expanded and presented in the upper panel of
Figure 3.
}
\end{figure*}

\begin{figure}
\epsfysize=14.5cm
\centerline{\epsfbox{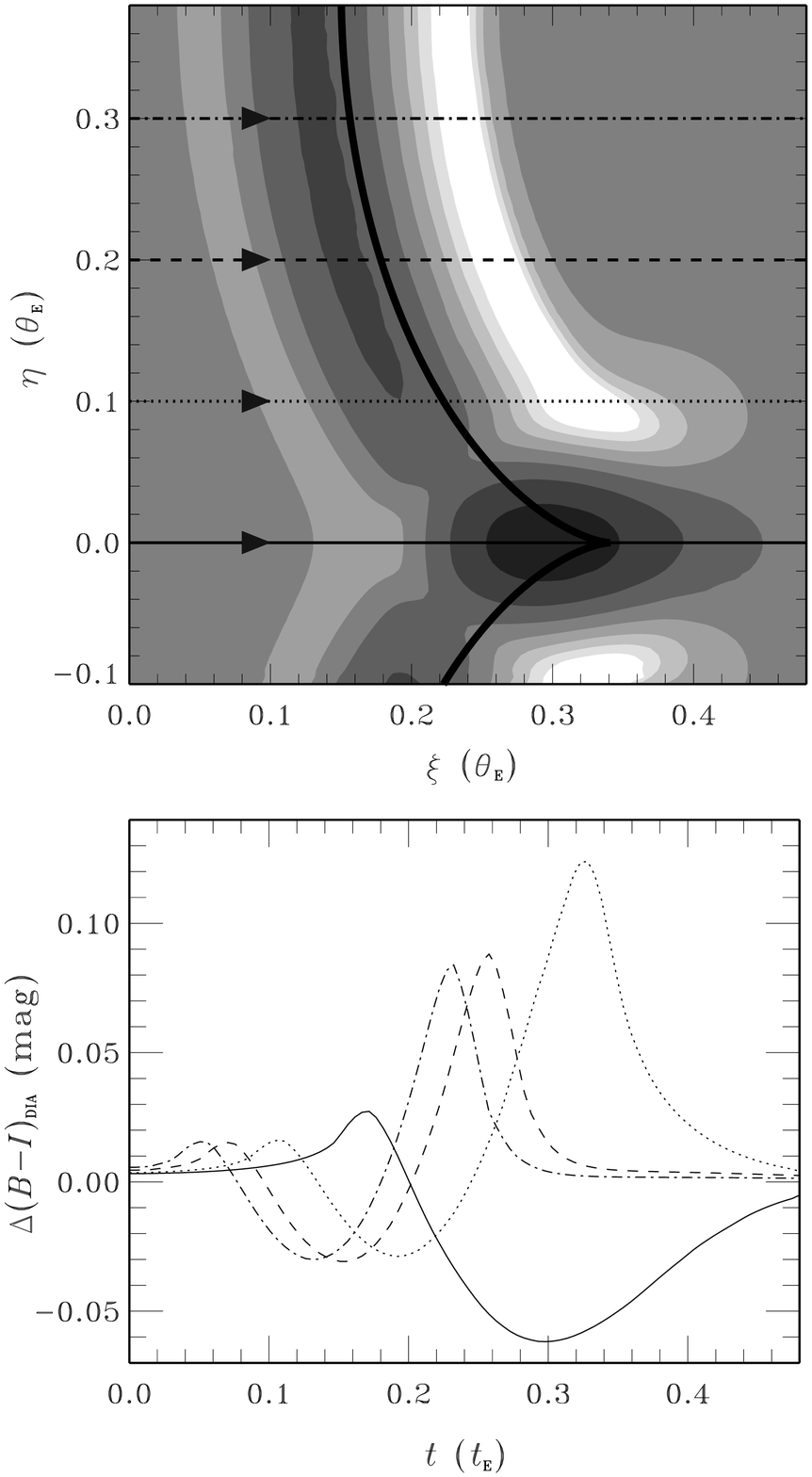}}
\caption{
The colour map (upper panel) near the caustic region and the colour curves
(lower panel) resulting from various source star trajectories (straight lines).
The region of the presented map corresponds to the enclosed part of the
map in Figure 2.  The grey tone of the map, the line types of the source
trajectories, and the corresponding colour curves are same as those in
Figure 2.
}
\end{figure}

From the colour map of the single lens system and the resulting colour curves,
one finds the following patterns.  First, the iso-colour-change contours
are concentric circles with their center at the center of the source star.
Due to the radial symmetry of the map, all the resulting colour curves
are symmetric with respect to the time of maximum amplification.  Second,
the colour does not monotonically change as a function of the separation
between the lens and the center of the source star, $r_L=\sqrt{x_L^2+y_L^2}$.
Outside the source star disk, the colour becomes redder as $r_L$ decreases,
but within the disk it becomes bluer with the decreasing value of $r_L$.
As a result, while the colour of the non-source transit event continues to
become redder as the lens approaches the source star, the colour curve of
a source transit event is characterized by turns at the moment when the
lens enters (or leaves) the source star surface.  Measurement of the turning
time, $t_\cap$, is important because one can determine the source radius
crossing time from the measured value of $t_\cap$ by
\begin{equation}
t_\ast=\left[\beta^2+ \left( {\vert t_\cap-t_0\vert 
\over t_{\rm E}} \right)^2\right]^{1/2} t_{\rm E},
\end{equation}
where the lensing parameters of $\beta$, $t_0$, and $t_{\rm E}$ are
determined from the overall shape of the light curve.  We note that the
described patterns of the DIA colour curves are very similar to those
of the un-blended event colour curves constructed by using the conventional
method (see Han, Park, \& Jeong\ 2000).

The patterns of the colour map for the binary lens system and the resulting
colour curves are as follows.  First, compared to the single lens map, the
the binary lens map is much more complex.  In the regions around fold
caustics, the iso-colour-change contours are parallel with the caustics.
However, we note that the contours on the left and right sides of the
caustics are not symmetric with respect to the caustic line.  In the regions
around cusps, on the other hand, the contours form separate peaks of blue
colour changes.  Second, the colour curves of caustic crossing binary lens
events are characterized by the turns during caustic crossings.  The colour
becomes redder as the source approaches the lens caustics.  It becomes
most reddish at around the time when the edge of the source star touches
the caustics.  As the source further approaches to the caustics and thus
the inner region of the source star disk lies on the caustics, the colour
becomes bluer.  During each caustic crossing, there are two red peaks which
occur when the left and right sides of the source star disk lie on the
caustics.  By measuring the time separation between the two consecutive red
peaks, $\Delta t$, one can estimate the approximate value of the source
radius crossing time by
\begin{equation}
t_\ast \sim {\Delta t\over 2 \sin \phi},
\end{equation}
where $\phi$ is the angle between the source trajectory and the tangential
line of the fold caustic at the position where the source crosses the
caustics.\footnote{If the fold caustic is a straight line, the relation
in equation (14) is exactly valid.  However, for an extended source event
the relation is an approximation because the fold caustics can no longer be
treated as a straight line.}  The value of $\phi$ can be determined from
the global fit of the binary lens event light curve.

\section{Uncertainties in DIA Colour Measurements}

In this section, we show by example that although the colour changes
induced by microlensing are usually small, they can be measured with
uncertainties small enough for one to extract useful information about
the lens and source stars.  We estimate the uncertainties of measured
colours, $\delta [\Delta(m_{\nu_2}-m_{\nu_1})_{\rm DIA}]$, in the
following ways.  Since the uncertainty in the determined source star
flux in magnitude is related to the signal-to-noise ratio by
\begin{equation}
\delta m_{\nu} = 
{(\delta F_{0,\nu}/F_{0,\nu}) \over 0.4\ {\rm ln}\ 10} 
\sim {1.09 \over S/N},
\end{equation}
the uncertainty in the measured colour is related to the signal-to-noise
ratio by
\begin{equation}
\delta [\Delta(m_{\nu_2}-m_{\nu_1})_{\rm DIA}]\sim 
\sqrt{2} \delta m_{\nu} \sim {1.54\over S/N}.
\end{equation}
Then, if $S/N \sim 10$, the uncertainty of the colour measurement will
be $\delta \Delta(m_{\nu_2}-m_{\nu_1})_{\rm DIA} \sim 0.15\ {\rm mag}$.
The signal measured from the subtracted image is proportional to the
source flux variation, i.e.\ $S\propto (A_\nu -1)F_{0,\nu} t_{\rm exp}$,
while the noise originates from both the lensed source and blended background stars,
i.e.\ $N\propto [(F_{0,\nu}A_\nu + \langle B\rangle)t_{\rm exp}]^{1/2}$.
Here $t_{\rm exp}$ is the mean exposure time and $\langle B\rangle$
represents the average total flux of faint unresolved stars within
the effective seeing disk with a radius $\theta_{\rm see}$.  Then
the signal-to-noise ratio is computed by
\begin{equation}
S/N = F_{0,\nu} (A_\nu -1) \left( {t_{\rm exp}\over F_{0,\nu}A_\nu
+ \langle B\rangle} \right)^{1/2}.
\end{equation}

By using equations (16) and (17), we estimate the uncertainties in the
measured colours for several example Galactic bulge single lens events
with various impact parameters.  We assume the events have a common
source size of $\theta_\ast=0.0756$ and an Einstein time scale of
$t_{\rm E}=67.5/2$ days by adopting the values of the MACHO Alert 95-30
for which the extended source effect was actually detected (Alcock et
al.\ 1997c).  The lensed source is a K-type giant with $I=14.05$ mag.
The observations are assumed to be carried by using a 1 m telescope with
a CCD camera that can detect 12 photons/s for $I=20$ star.  Note that the
detection rate is determined considering both the efficiency and gain of
the CCD camera.  Since the major targets for colour measurements are high
amplification events with bright source stars, proper exposure time is
important to prevent saturation of images.  In addition, since the
amplification varies rapidly for these events, we leave $t_{\rm exp}$ as
a variable so that the measured signal to be $\sim 40,000$ photons,
which corresponds to the median flux of the linear sensitivity region
for typical modern CCD cameras.
We estimate $\langle B\rangle$ by assuming that blended light comes
from stars fainter than the crowding limit.  Towards the Galactic bulge
field, the crowding limit is set when the stellar number density reaches
$\sim 10^6$ stars/${\rm deg}^2$.  Based on the model luminosity function
constructed by combining those of Holtzman et al.\ (1998) and Gould,
Bahcall, \& Flynn (1997)\footnote{For the detailed method of the model
LF construction, see Han, Jeong, \& Kim (1998).}, this number density
corresponds to the de-reddened $I$-band magnitude of $I_0\sim 18.1$.  The
background flux is normalized for stars in the seeing disk with
$\theta_{\rm eff}=2''\hskip-2pt .0$.

\begin{figure}
\vskip-0.6cm
\epsfysize=9.0cm
\centerline{\epsfbox{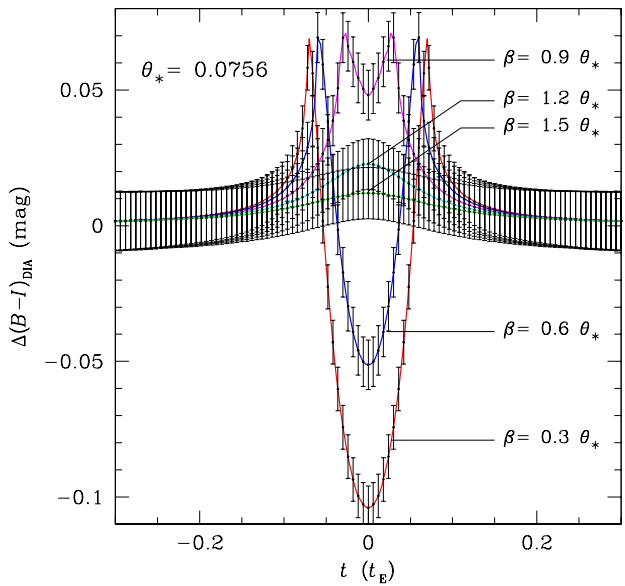}}
\vskip-0.5cm
\caption{
The colour curves of several example microlensing events along with the
estimated uncertainties, which are presented by error bars. We assume the
events have a common source size of $\theta_\ast = 0.0756$.  For the
details of error estimates, see \S\ 5.
}
\end{figure}

In Figure 4, we present the colour curves of the events along with the
estimated uncertainties, which are presented by error bars.  From the
figure, one finds that it will be difficult to distinguish colour change
curves for different non-source transit events ($\beta > \theta_\ast$)
due to large uncertainties.  However, the uncertainties for source transit
events ($\beta\leq\theta_\ast$) are small enough for one to distinguish
curves with different values of $\beta$ and determine the turning times.

Although it is not included in the above uncertainty estimate, another
possible source of uncertainty comes from the residual noise due to
flat fielding (C.\ Alard, private communication).  Since this noise
is constant, it becomes important especially in the wings of the colour
change curve at which the signal is very low as $A_\nu-1 \rightarrow 0$.
However, we find that in the region near the peak amplification where
most information from the colour curve is obtained, this noise
is not important.  For example, let us the residual noise $\delta F_\nu$
to be $\sim 5\%$ of the unlensed source star flux, i.e.\ $\delta F_\nu 
\sim 0.05 F_{0,\nu}$.  In the region close to the source star, e.g.\
$\beta \lesssim 0.1$ (and thus $A\gtrsim 10$), the uncertainty in the
DIA colour measurements estimated by using equations (15) and (16) will
be $\lesssim 0.008$ mag for the same source of the example event in the
above error estimate.  This uncertainty is substantially (nearly an order)
smaller than the colour change of $\gtrsim 0.05$ mag for source transit
events.

\section{Summary}

We investigate the chrometicity of single and binary microlensing events
when their colour changes are measured by using the recently developed
DIA photometric method.  The findings from this investigation are
summarized as follows:

\begin{enumerate}
\item
The microlensing induced colour change measured by the DIA method differs
from that measured by using the conventional photometric method.

\item
Despite the difference, from the DIA colour curve one can obtain the
same information about the lens and source star as the information
obtained from the conventional colour curve.  This is because both
colour curves depend on the same parameters.

\item
However, since the DIA colour curve is not affected by blending, one
can determine the lens proper motion and source star surface intensity
profile with significantly reduced uncertainties.

\item
While the colour of a non-source transit event continues to become
redder as the lens approaches the source star, the colour curve of a
source transit event is characterized by turns the moments when the
lens enters and leaves the source star disk.  For the source transit
event, one can determine the source radius crossing time by measuring
the turning time of the colour curve.

\item
The colour curveas of binary lens events are much less symmetric and
vary greatly depending on the lens system geometry and the source star
trajectory.  For a caustic crossing event, one can measure the source
radius crossing time by measuring the time separation between the two
consecutive red peaks on the colour curve.
\end{enumerate}

This work was supported by the grant (KRF-99-041-D00442) of the Korea
Research Foundation.

\end{document}